\documentclass{ws-mpla}
\usepackage[super]{cite}
\usepackage{graphicx}
\usepackage{hyperref}
\usepackage{url}
\usepackage{xcolor}

\begin{document}

\markboth{Saleev V. A., Shilyaev K. K.}{Production of $S$-wave
charmonia in the SGR approach using the NRQCD}

\catchline{}{}{}{}{}

\title{Production of $\bm{S}$-wave charmonia\\ in the soft gluon resummation approach using the NRQCD}

\author{Vladimir Saleev}

\address{Samara National Research University, Moskovskoe Shosse,
34, 443086, Samara, Russia }
\address{Joint Institute for Nuclear Research, Dubna, 141980
Russia.\\
saleev.vladimir@gmail.com}

\author{Kirill Shilyaev}

\address{Samara National Research University, Moskovskoe Shosse,
34, 443086, Samara, Russia. \\
kirillsept@gmail.com
 }

\maketitle

\pub{Received (Day Month Year)}{Revised (Day Month Year)}

\begin{abstract}
In this article, we study the $S$-wave charmonium production at the small
transverse momenta in the transverse momentum dependent (TMD) parton
model (PM), as it is formulated in the soft gluon resummation (SGR)
approach, using the nonrelativistic quantum chromodynamics (NRQCD) as
the model of the heavy quarkonium hadronization. To extend the scope
of the calculation to the phenomenologically important region of
intermediate transverse momenta, we use the inverse-error weighting
(InEW) matching scheme to combine the TMD PM predictions with the
results of the collinear PM (CPM) fixed-order calculations, which
are applicable at the large transverse momenta. We predict the
direct $\eta_c$ and the prompt $J/\psi$ production cross sections
and transverse momentum spectra in proton-proton collisions at the
LHC, RHIC, AFTER and NICA energies. Predictions for the polarized
$J/\psi$ production processes are also presented.

\end{abstract}


\ccode{PACS Nos.: 13.60.Le; 13.88.+e.}

\section{Introduction}
\label{Introduction}

Description of the heavy quarkonium production of mass $M$ in the
domain of small transverse momenta, i.e. $p_T^{} \ll M$, is mostly
more complicated than the   one at the large transverse momenta,
$p_T^{} \gg M$, where the collinear parton model (CPM) including the
fixed-order perturbative calculation in the quantum chromodynamics
(QCD) is the conventional approach \cite{Chapon:2020heu}.  The most
adequate approach for the small-$p_T$ kinematical region is the
transverse momentum dependent (TMD) parton model
(PM)~\cite{Collins:2011zzd, Boussarie:2023izj}. In the TMD PM, the
soft-hard factorization theorem is originally applied to describe
the small-$p_T$ production of the Drell-Yan pairs, $W$ and $Z$
bosons \cite{Collins:1984kg,Moos:2023yfa, Bacchetta:2018lna,
Bacchetta:2019sam, BermudezMartinez:2020tys}, and Higgs bosons
\cite{Kauffman:1991jt,Gutierrez-Reyes:2019rug, Boer:2011kf,
Boer:2014lka}, i.e. colorless particles, for which both the final
state gluon radiation and the final state soft gluon exchange are
absent. In this case, the soft gluon initial radiation and the soft
gluon exchange can be factorized in the TMD parton distribution
functions (PDFs). The latter ones include dependencies on the
initial longitudinal and transverse momenta of the partons and these
dependencies are not split up in general. The evolution of TMD PDFs
with a factorization scale $\mu$ and a rapidity separation scale
$\zeta$ is controlled by the system of the Collins-Soper-Sterman
(CSS) differential equations~\cite{Collins:2011zzd,Collins:1988ig}.


Here we study the charmonium production via gluon fusion
subprocesses in collisions of unpolarized protons. In general, the
unpolarized and linearly polarized gluons may contribute to the
production cross section~\cite{Boer:2016bfj}. {The contribution in
the transverse momentum spectrum of a linearly polarized gluon TMD
PDF, the so-called Boer-Mulders PDF, is estimated to be relatively
small (10-20\%), when it is calculated using the SGR approach
{\cite{Boer:2016bfj}}. Such a way, we omit this contribution during
the presented study.}

In order to describe the heavy quark-antiquark pair hadronization
into a final heavy quarkonium we use the model of nonrelativistic
quantum  {chromodynamics} (NRQCD)~\cite{Bodwin:1994jh}.  The NRQCD
includes contributions of color singlet and color octet intermediate
states, for the latter ones the soft gluon final state radiation
can't be ignored. Formally, TMD factorization can be used but TMD
PDFs become process- {dependent}. As recently shown in
Ref.\cite{Echevarria:2019ynx}, the proper TMD factorization for
quarkonia production requires the introduction of a new
non-perturbative hadronic quantities beyond the TMD PDFs: the TMD
shape functions. It seems that the use of heavy quarkonium
production to probe gluon TMD PDFs is still a very complicated task.

Taking in mind the phenomenological aim of our study, to provide
predictions of total cross sections and transverse momentum spectra
of $S$-wave charmonia relevant for future experiments, we use here
 {the} well-known soft gluon resummation (SGR)
approach~\cite{Collins:1981va,Collins:1984kg} as a realization of
the TMD factorization for modelling of the
 {non-collinear} parton dynamics at small transverse
momentum  {of charmonium}. In general, our approach is close to the
recent studies in Refs.~\cite{Boer:2014tka, Sun:2012vc}.

In order to extend the scope of the calculation to the phenomenologically
important region of intermediate transverse momenta, we use the
inverse-error weighting (InEW) matching
scheme~\cite{Echevarria:2018qyi} to combine the TMD PM predictions
and the results of the collinear PM (CPM) fixed-order calculations,
which are applicable at the large transverse momenta.

\section{Soft gluon resummation approach}
\label{SGRA}

The 4-momenta of the initial partons within the TMD PM are  {taken
on-shell} $q_{1,2}^2 = 0$ and can be written in the Sudakov
decomposition form:
\begin{equation}
q_{1}^{\mu} = x_1^{} p_1^{\mu} + y_1 p_2^{\mu} + q_{1T}^{\mu}, ~~~~~
q_{2}^{\mu} = x_2^{} p_2^{\mu} + y_2 p_1^{\mu} + q_{2T}^{\mu},
\end{equation}
where $p_{1,2}^{} = \frac{\sqrt{s}}{2} (1, 0, 0, \pm 1)$ are
4-momenta of colliding protons, $x_{i}^{}$ and $y_{i}^{} =
\vec{q}_{iT}^{\,\,2} / (s x_i^{}) $ are longitudinal fractions of
the parton momenta and ${\vec q}_{iT}^{}$ are their transverse momenta
($q_{iT}^2 = - \vec{q}_{iT}^{\,\,2} $). The TMD factorization
prescribes the initial transverse momenta to be small compared to
the hard scale of the process $\mu$  which is of order of charmonium
mass $M$, so that the corrections of order $\mathcal{O}
(\vec{q}_{iT}^{\,\,2} / M^2)$ are neglected here, and therefore
$y_{1,2} \approx 0$. Then the initial 4-momenta are written as
follows:
\begin{equation}
q_1^{} \approx \left( \frac{x_1^{} \sqrt{s}}{2}, \vec{q}_{1T}^{},
\frac{x_1^{} \sqrt{s}}{2} \right), ~~~~~ q_2^{} \approx \left(
\frac{x_2^{} \sqrt{s}}{2}, \vec{q}_{2T}^{}, -\frac{x_2^{}
\sqrt{s}}{2} \right).
\end{equation}

The TMD PM allows, as mentioned above, to describe a production
cross section as a convolution of a partonic $2\to 1$ cross section
$d\hat{\sigma}$ and TMD PDFs~\cite{Collins:2011zzd}:
\begin{equation}
d\sigma^{TMD} = \int d x_1^{}\, d x_2^{}\, d^2 q_{1T}^{}\, d^2
q_{2T}^{}\, F(x_1, \vec{q}_{1T}, \mu, \zeta_1^{})\, F(x_2,
\vec{q}_{2T}, \mu, \zeta_2^{})\, d\hat{\sigma}^{2\to 1}.
\label{eq:factorization}
\end{equation}
The $2 \to 1$ subprocesses of the charmonium $\cal C$ production
with quarks and gluons $(gg\to {\cal C}, q\bar q\to {\cal C})$ are
taken into account because they are at leading order at small~$p_T^{}$, the cross section of the corresponding subprocess is
\begin{equation}
d\hat{\sigma}^{2\to 1} = (2\pi)^4 \delta^{(4)} (q_1^{} + q_2^{} - p)
\frac{\overline{|\mathcal{M}(2 \to 1)|^2}}{I} \frac{d^3 p}{(2\pi)^3
2p_0}\label{eq:21}
\end{equation}
where $I \approx 2 x_1^{} x_2^{} s$ is the flux factor and
$\mathcal{M}(2 \rightarrow 1)$ is the amplitude of a gluon-gluon
fusion subprocess $gg\to {\cal C}$ or a quark-antiquark annihilation
subprocess $q \bar q \to {\cal C}$, and ${\cal C}=\eta_c, J/\psi,
\chi_{cJ}, \psi'$. Taking in mind the condition $p_T \ll M$, we can
rewrite (\ref{eq:21}) as follows:
\begin{equation}
\frac{d\hat \sigma^{2\to 1}}{ dp_T dy}=\frac{2\pi^2 p_T}{s
M^2}\overline{|\mathcal{M}(2 \rightarrow 1)|^2}
\delta^{(2)}\left({\vec q}_1+{\vec q}_2-{\vec p_T}\right)\delta
\left(x_1-\frac{M e^y}{\sqrt{s}}\right) \delta\left(x_2-\frac{M
e^{-y}}{\sqrt{s}}\right) .
\end{equation}
In this way, the factorization formula (\ref{eq:factorization}) looks
like this:
\begin{equation}
\frac{d\sigma^{TMD}}{dp_T dy}= \frac{2\pi^2 p_T}{s M^2}
\overline{|\mathcal{M}(2 \to 1)|^2}  F(x_1, \vec{q}_{1T}, \mu,
\zeta) \otimes_T  F(x_2, \vec{q}_{2T}, \mu,
\zeta),\label{eq:factorizaion2}
\end{equation}
where $y$ is a rapidity of the final quarkonium and
$$ F(x_1,
\vec{q}_{1T}) \otimes_T  F(x_2, \vec{q}_{2T}) = \int d^2q_{1T} d^2
q_{2T} F(x_1, \vec{q}_{1T})F(x_2, \vec{q}_{2T}) \delta^{(2)}({\vec
q}_1+{\vec q}_2-{\vec p_T}).$$

The  evolution of the TMD PDF with respect to the scales should be
implemented after the two-dimensional Fourier transform of the
PDF~\cite{Collins:2011zzd}:
\begin{equation}
\hat{F} (x, \vec{b}_{T}^{}, \mu, \zeta) = \int d^2 q_{T}^{}\, e^{i
\vec{q}_{T}^{}\,
 \vec{b}_{T}^{}} F(x, \vec{q}_{T}^{}, \mu, \zeta).
\end{equation}
In terms of the  Fourier-conjugated TMD PDFs, we can rewrite the formula
(\ref{eq:factorizaion2}):
\begin{equation}
\frac{d\sigma^{TMD}}{dp_T dy}= \frac{p_T}{2 s M^2}
\overline{|\mathcal{M}(2 \to 1)|^2} \int d^2 b_T {e}^{-i \vec{b}_{T}
\vec{p}_{T}}  {\hat F}(x_1, \vec{b}_{T}, \mu, \zeta) {\hat F}(x_2,
\vec{b}_{T}, \mu, \zeta) \label{eq:factorizaion3}.
\end{equation}

Within the SGR approach, the perturbative evolution of the
Fourier-conjugated PDFs in an impact parameter $\vec{b}_{T}^{}$
space is implemented in a multiplication way~\cite{Collins:1981uk}:
\begin{equation}
\hat{F}(x_1^{}, b_{T}^{}, \mu, \zeta) \hat{F}(x_2^{}, b_{T}^{}, \mu, \zeta)
 = e^{-S_P^{} (\mu, \mu_{b}^{}, b_{T}^{})} \hat{F}(x_1^{}, b_{T}^{}, \mu_{b}^{},
  \mu_{b}^{2}) \hat{F}(x_2^{}, b_{T}^{}, \mu_{b}^{},
  \mu_{b}^{2})\label{eq:sgr1}
\end{equation}
where the standard simple choice for the scales $\mu^{} =
\sqrt{\zeta}$ was done, and the scale $\mu_0=  \sqrt{\zeta_0^{}} =
\mu_b$  {was} taken as the initial one, the expression for which is
given below. This choice is made to minimize large values of the
logarithms of the scale ratios $\mu/\Lambda_\textnormal{\scriptsize
QCD}^{}$, $\mu/M$~\cite{Bor:2022fga}.

In equation (\ref{eq:sgr1}), the function $S_P^{}$ is  {the}
so-called Sudakov factor which realizes the evolution from the
initial scales $(\mu_0=\mu_b, \zeta_0=\mu_b^2)$ to the final ones
$(\mu, \zeta=\mu^2)$. The Sudakov factor is written this
way~\cite{Collins:1981uk, Boer:2014tka}:
\begin{equation}
S_P^{} (\mu, \mu_b^{},b_T) =  \int\limits_{\mu_b^2}^{\mu^2}
\frac{d\mu'^{2}}{\mu'^{2}} \left[ A(\mu') \ln \frac{\mu^2}{\mu'^2} +
B(\mu') \right],
\end{equation}
where the coefficients $A$ and $B$ are presented as a series with
respect to the strong coupling constant, and their terms represent
logarithmic orders of calculation via $A^{(n)}$ coefficients and
orders with respect to the $\alpha_s$ via $B^{(n)}$:
\begin{equation}
A(\mu') = \sum\limits_{n=1}^{\infty} A^{(n)} \left( \frac{\alpha_s(\mu')}{\pi} \right)^n,
 ~~~~~ B(\mu') = \sum\limits_{n=1}^{\infty} B^{(n)} \left( \frac{\alpha_s(\mu')}{\pi} \right)^n,
\end{equation}
and the leading logarithmic (LL) and LO in the $\alpha_s$
approximation of the SGR approach (LL-LO) corresponds  {to the}
first coefficients of the series
\begin{align}
A^{(1)} &= C_A, \\
B^{(1)} &= - \frac{11 C_A - 2 N_f}{6} - \frac{C_A}{2} \delta_{c8},
\end{align}
the quantity $\delta_{c8}$ is equal to $0$ for color singlet states
and $1$ for color octet quarkonium states, $N_f$ is a number of
quarks flavors, $N_c = 3$ is a number of colors and $C_A =N_c = 3$.

In the one-loop approximation for the coupling constant
$\alpha_s^{}$, an explicit analytical expression for the integral in
the function $S_P^{}$ can be obtained~\cite{Bor:2022fga}. The
expression for the Sudakov factor $S_P^{}$ is applicable in the
range $b_0/Q \leqslant b_T^{} \leqslant b_{T,\,
\textnormal{\scriptsize max}}^{}$, where $b_0 = 2 e^{-\gamma}$,
$\gamma$ is the Euler--Mascheroni constant. The lower limit of the
range is given by the expression $\mu_b\to \mu_b'^{} = Q b_0/ (Q
b_T^{} + b_0)$, and the upper limit~\cite{Collins:1984kg} is
determined by replacing the impact parameter with $b_T \to  b_T^{*}
(b_T^{}) = b_T^{}/\sqrt{1+(b_T^{}/b_{T,\,\textnormal{\scriptsize
max}})^2}$. We used the largest value of the impact parameter
$b_{T,\,\textnormal{\scriptsize max}} = 1.5$ GeV$^{-1}$. There is
some freedom of choice in both the $b_{T,\,\textnormal{\scriptsize
max}}$ parameter and in the form of the $b_T^{}$ prescription; other
possible choices and their reasonings can be found in
Ref.~\cite{Bor:2022fga}.

In addition, the suppression of the $S_P^{}$ at large $b_T^{}$ is
guaranteed by the nonperturbative Sudakov factor $S_{NP}^{}$, the
expression for which is not theoretically derived, so that the function
$S_{NP}^{}$ is extracted from experimental data. In the
Ref.~\cite{Aybat:2011zv}, the parameterization in a Gaussian form
was obtained for the quark-antiquark annihilation processes:
\begin{equation}
\tilde S_{NP}^{} (x, b_T^{}, \mu) = \frac{1}{2} \left[ g_1 \ln
\frac{\mu}{2Q_{NP}} + g_2 \left( 1 + 2 g_3 \ln \frac{10 x
x_0^{}}{x_0 + x} \right) \right] b_T^2\label{eq:sudakov_NP}
\end{equation}
with $g_1=0.184$ GeV$^2$, $g_2=0.201$ GeV$^2$, $g_3=-0.129$,
$x_0=0.009$, $Q_{NP}=1.6$~GeV. Due to the lack of experimental data,
it is also applied for gluon-gluon fusion processes but with an
additional color factor change $C_A^{}/C_F^{}$~\cite{Bor:2022fga},
where $C_F=(N_c^2-1)/(2 N_c)=4/3$. As two PDFs are included in a
cross section expression, we can write the nonperturbative Sudakov
factor for the product of the PDFs in this way:
\begin{equation}
S_{NP}^{}(x_1,x_2,b_T, \mu) = \tilde S_{NP}^{}(x_1,b_T,\mu) + \tilde
S_{NP}^{}(x_2,b_T,\mu).
\end{equation}

In the SGR approach, the TMD PDFs are expressed with collinear
parton distributions at the initial scale $\mu_b'^{}$ and at the
leading order  {the} one can be written as follows:
\begin{equation}
\hat{F}(x, b_T^{},\mu_b'^{},{\mu_b'}^{2} ) = f (x, \mu'^{}_{b}) +
\mathcal{O} (\alpha_s) + \mathcal{O} (b_T^{}
\Lambda_\textnormal{\scriptsize QCD}).
\end{equation}
They lack the   genuine, non-perturbative content, which is a
subject of different non-perturbative models of TMD PDFs at the low
hard scale, see Refs.
\cite{Bacchetta:2020vty,Chakrabarti:2023djs,Bacchetta:2024fci,
Yu:2024mxo}. However, such information is included in the SGR
calculations via the non-perturbative Sudakov factor $S_{NP}$
(\ref{eq:sudakov_NP}). As it will be demonstrated below in Sec.
\ref{results}, our predictions for $\eta_{c}^{}$ production cross
sections at the energies 115 GeV and 7 TeV based on the SGR approach
  coincide with  {the} results obtained
{earlier} in Ref.\cite{Bacchetta:2022nyv} using the spectator model
for non-perturbative gluon TMD PDF,  with accuracy about 10-20~\%,
see Figs. \ref{fig:1} and \ref{fig:2}.

Then the master formula for calculation of the differential cross section
in the SGR approach reads
\begin{equation}
\frac{d^2\sigma}{dp_T^{} dy} = \frac{\pi p_T^{}
\overline{|\mathcal{M}
 (2 \rightarrow 1)|^2}}{M^2 s}
 \int d b_T^{}\,b_T^{}\, J_0(p_T^{} b_T^{})\,\,e^{-S_P}\, e^{-S_{NP}} \, {f (x_1, \mu'^{}_{b*})\, f (x_2,
 \mu'^{}_{b*})},
\end{equation}
where $S_P=S_P^{}(\mu, \mu_{b*}',b_T^{*})$ and
$S_{NP}=S_{NP}^{}(x_1,x_2,\mu,b_T^{})$, $J_0$ is the first kind
Bessel function of the zeroth order.

\section{Collinear parton model}
\label{cpm}

The relevant LO in the strong coupling constant $\alpha_s$
calculation within the CPM based on the CPM factorization theorem:
\begin{equation}
d\sigma^{CPM} = \int d x_1^{}\, \int d x_2^{}\, f(x_1^{}, \mu)\,
f(x_2^{}, \mu)\, d\hat{\sigma} (s, x_1^{}, x_2^{}),\label{eq:cpm}
\end{equation}
where $f(x_{1,2}^{}, \mu)$ are gluon (quark and antiquark) collinear
PDFs. The partonic cross section is written in the standard way for
the $2 \rightarrow 2$ subprocesses:
\begin{equation}
d\hat{\sigma} = (2\pi)^4 \delta^{(4)} (q_1^{} + q_2^{} - p - k)
\frac{\overline{|\mathcal{M}(2 \to 2)|^2}}{I} \frac{d^3 p}{(2\pi)^3
2p_0} \frac{d^3 k}{(2\pi)^3 2k_0}.\label{eq:cpm2}
\end{equation}
Combining (\ref{eq:cpm}) and (\ref{eq:cpm2}) together, we find the
basic formula for the numerical calculation in the CPM:
\begin{equation}
\frac{d\sigma^{CPM}}{dy dp_T}=\frac{p_T}{8 \pi s}\int
\frac{dx_1}{x_1} \int \frac{dx_2}{x_2}\, f(x_{1}^{}, \mu)\,
f(x_{2}^{}, \mu)\, \overline{|\mathcal{M}(2 \to 2)|^2} \delta \left(
\hat s+\hat t +\hat u - M^2 \right),
\end{equation}
where $\hat s=x_1 x_2 s$, $\hat t=M^2-\sqrt{s} M_T e^{-y}$, $\hat
u=M^2-\sqrt{s} M_T e^{y}$, and $M_T=\sqrt{M^2+p_T^2}$. Considering only the LO contributions from gluon-gluon fusion and
quark-antiquark annihilation partonic subrocesses, we included in
the LO CPM calculation the next ones: $gg\to {\cal C} g$ and $q \bar
q\to {\cal C}g$, where ${\cal C}=J/\psi, \eta_c, \chi_{cJ}, \psi'$.

In the LO CPM calculations, as well as in the calculations based on
the SGR approach, to describe the prompt $J/\psi$ production data, we
calculate the sum of the direct production cross section and the feed-down
contributions from decays $\chi_{cJ}\to J/\psi \gamma$ and $\psi'
\to J/\psi X$.

\section{Nonrelativistic QCD}
\label{nrqcd}

 The nonrelativistic QCD (NRQCD) is a conventional
approach for the description of the hadronization of heavy
quark-antiquark pair into heavy quarkonium~\cite{Bodwin:1994jh}. The
large mass of the charm quarks $m_c^{}$ allows us to consider them
as { nonrelativistic ($\upsilon^2 \approx 0.3$)~\cite{Lucha:1991vn,
Eichten:1995ch}}, and therefore the following dynamical observables:
quarkonium mass, momentum, kinetic energy, etc., are quite reliably
separated by orders of their magnitudes~\cite{Lepage:1992tx}. The
estimation of the magnitude of observables gives the right to
introduce a hierarchy of Fock states of charmonium in the $J/\psi$
production with respect to the relative velocity $\upsilon$ of the
constituent quarks~\cite{Cho:1995vh}:
\begin{multline}
|J/\psi\rangle = \mathcal{O}(\upsilon^0) |c\bar{c}[{}^3S^{(1)}_1]\rangle + \mathcal{O}(\upsilon^1) |c\bar{c}[{}^3P^{(8)}_J]g\rangle + \\ + \mathcal{O}(\upsilon^2) |c\bar{c}[{}^3S^{(1,8)}_1]gg\rangle + \mathcal{O}(\upsilon^2) |c\bar{c}[{}^1S^{(8)}_0]g\rangle + \dots
\end{multline}
The corresponding expansion for the $\eta_c$ state looks as follows:
\begin{multline}
|\eta_c\rangle = \mathcal{O}(\upsilon^0) |c\bar{c}[{}^1S^{(1)}_0]\rangle + \mathcal{O}(\upsilon^1) |c\bar{c}[{}^1P^{(8)}_1]g\rangle + \\ + \mathcal{O}(\upsilon^2) |c\bar{c}[{}^3S^{(1,8)}_1]g\rangle + \mathcal{O}(\upsilon^2) |c\bar{c}[{}^1S^{(1,8)}_0]gg\rangle + \dots
\end{multline}
and for $P$-wave charmonia $\chi_{cJ}$ it is
\begin{equation}
|\chi_{cJ}\rangle = \mathcal{O}(\upsilon^0) |c\bar{c}[{}^3P^{(1)}_J]\rangle + \mathcal{O}(\upsilon^1) |c\bar{c}[{}^3S^{(8)}_1]g\rangle + \dots
\end{equation}

The leading term of the series is the color singlet Fock state, in
which the constituent quarks are in the observable charmonium. If we
keep only this term, then this approximation is called the Color
Singlet Model (CSM)~\cite{Kuhn:1979bb}.

In the NRQCD, the production cross section of a charmonium state is
factorized into the cross section of a quark-antiquark pair
production in some Fock state and a long-distance matrix element
(LDME), which can be interpreted as describing the hadronization of
a quark-antiquark pair into a bound state:
\begin{equation}
d\hat{\sigma} ( a + b \rightarrow \mathcal{C} + X ) = \sum\limits_n d\hat{\sigma} (a + b \rightarrow c\bar{c}[n] + X ) \langle \mathcal{O}^{\mathcal{C}}[n] \rangle / (N_\textnormal{\scriptsize col}^{} N_\textnormal{\scriptsize pol}^{}),
\end{equation}
where the short notation $n={}^{2S+1}L_{J}^{(1,8)}$ denotes the Fock
state in the color singlet~$^{(1)}$ or color octet~$^{(8)}$ state,
with spin $S$, orbital momentum $L$, total angular momentum $J$,
$N_\textnormal{\scriptsize pol}^{} = 2J+1$,
$N_\textnormal{\scriptsize col}^{} = 2N_c^{}$ for color singlets,
$N_\textnormal{\scriptsize col}^{} = N_c^2 - 1$ for color octets,
and $N_c^{}=3$. The amplitude of the quark-antiquark pair production
in the respective Fock state is calculated in the fixed-order of the
perturbative QCD using the Feynman diagram technique and a sequence
of projections onto states with the necessary values of the quantum
numbers~\cite{Bodwin:1994jh,Cho:1995vh}.

In the SGR approach, we take into account the production of states
$^1S_0^{(8)}$, $^3P_{0,2}^{(8)}$ for $J/\psi$; $^3P_{0,2}^{(1)}$,
$^3S_{1}^{(8)}$ for $\chi_{cJ}^{}$; $^1S_0^{(1)}$ for $\eta_c$ (as
the CSM approximation is sufficient and color octet states lead to an
overestimation of the experimental data) in subprocesses $2 \to 1$. In
the CPM, the charmonia in $2 \to 2$ partonic subprocesses are
produced via the states $^3S_1^{(1)}$, $^3S_1^{(8)}$, $^1S_0^{(8)}$
and $^3P_J^{(8)}$ for $J/\psi$, $\psi'$ and via $^3P_J^{(1)}$,
$^3S_1^{(8)}$ for $\chi_{cJ}^{}$ . The hard squared matrix elements
of heavy quarkonium production in the gluon-gluon fusion and
quark-antiquark annihilation corresponding to the color singlet
states can be found in Ref.~\cite{Gastmans:1986qv}, and those for the
color octet states are in Ref.~\cite{Cho:1995ce}.

Nowadays, all known sets of LDME values are phenomenological. The
expressions for the color singlet LDMEs are related to the values of
the charmonium wave function or its derivative at the
origin~\cite{Lucha:1991vn, Eichten:1995ch}. These values are
obtained in calculations in nonrelativistic potential model with
phenomenological potentials or from experimental data for charmonium
decays. The LDME for color octet states can't be calculated in the
QCD and therefore they are extracted by fitting charmonium data
after subtraction of the color singlet contributions. Although the
LDME values are assumed to be universal, the results of LDME fits of
different data sets with various $\sqrt{s}$ and, especially, in
different orders of $\alpha_s$ calculations can differ quite
significantly. Therefore, in this study we obtain our own results
for octet LDMEs by the fitting of prompt $J/\psi$ production data at
$\sqrt{s} = 200$
GeV~\cite{PHENIX:2011gyb,PHENIX:2006aub,STAR:2009irl,PHENIX:2009ghc}.

{The $\eta_c$ production cross sections in proton-proton collisions
are measured only in experiments of the LHCb
collaboration~\cite{LHCb:2014oii,LHCb:2019zaj} at the $\sqrt{s}=7$
and $8$ TeV, and at the region of applicability of the CPM,
$6.5<p_T^{\eta_c}<14$ GeV.  As it was shown in Ref.
\cite{Butenschoen:2014dra}, the CPM using the CSM calculation may be
sufficient for a correct and complete description of the $\eta_c$
production cross section and $p_T$-spectrum. The similar results
have been obtained   in the calculations based on the high-energy
factorization approach and the CSM, see
Refs.~\cite{Babiarz:2019mag,Anufriev:2024rey}.}

\section{Matching scheme}
\label{matching}

For the region of intermediate transverse momenta $p_T^{} \sim M$,
there is no approach based on the perturbative expansion of the
cross section in a series and its representation in the factorized
form, as it is done by the collinear and TMD factorization theorems.
Instead, the contributions of the two factorization approaches can
be matched and, in this way, can describe the intermediate region of
$p_T^{}$ as a certain sum of the CPM and TMD PM contributions. We
use the Inverse-Error Weighting (InEW) scheme, which is based on the
inverse-variance weighting scheme for the evaluation of the weighted
average, such a method for choice of weights for random variables that the variance of their weighted sum is the
smallest~\cite{Echevarria:2018qyi}.

In the InEW scheme, the matched $\overline{d\sigma}$ cross section
is represented as the sum of the TMD PM contribution ${TMD}$ and the
collinear fixed-order term ${CPM}$ with the weights $\omega_1$ and
$\omega_2$:
\begin{equation}
\overline{d\sigma}(p_T^{},Q) = \omega_1 \, d\sigma^{TMD}(p_T^{},Q) +
\omega_2 \, d\sigma^{CPM}(p_T^{},Q).
\end{equation}
The normalized values of the inverse squares of the power
corrections used in the CPM and TMD PM are taken as  {the} weights:
\begin{equation}
\omega_1 = \frac{\Delta^{-2}_{{TMD}}}{\Delta^{-2}_{{TMD}} +
\Delta^{-2}_{{CPM}}},~~~~~~~ \omega_2 =
\frac{\Delta^{-2}_{{CPM}}}{\Delta^{-2}_{{TMD}} +
\Delta^{-2}_{{CPM}}},
\end{equation}
\begin{equation}
\Delta_{{TMD}}^{} = \left( \frac{p_T^{}}{Q} \right)^2 + \left(
\frac{m}{Q} \right)^2,~~~~~~~ \Delta_{{CPM}}^{} = \left(
\frac{m}{p_T^{}} \right)^2 \cdot \left( 1 + \ln^2 \left(
\frac{Q_T^{}}{p_T^{}}  \right)   \right),
\end{equation}
where $m$ is a hadronic mass scale of order $1$ GeV, $Q_T^{} =
\sqrt{Q^2 + p_T^2}$. During our calculations $Q=M$ is the charmonium
mass. The uncertainty of the cross section evaluation, defined as
the weighted average of the TMD PM and CPM contributions, is given
by the following expression:
\begin{equation}
\Delta \overline{d\sigma} = \frac{ \Delta_{{TMD}}^{}
\Delta_{{CPM}}^{}}{\sqrt{\Delta_{{TMD}}^{2} + \Delta_{{CPM}}^{2}}}
\overline{d\sigma}.
\end{equation}

Thus, the InEW scheme allows us to calculate a cross section that is
reduced to the contributions of the CPM and TMD PM in their regions
of applicability and which is represented as a weighted average of
these contributions in the region where neither theorem is strictly
applicable. The error of the final cross section is found to be of
the maximum value in the region of intermediate transverse momenta.

\section{Results}
\label{results}

All numerical calculations described below were performed using the
numerical integrator CUBA~\cite{Hahn:2004fe} with a maximum relative
error of $1\%$. The collinear PDFs were taken as numerically defined
distributions MSTW2008LO~\cite{Martin:2009iq}.

The masses~\cite{ParticleDataGroup:2020ssz} of the charmonium states
used in the calculations are $M_{J/\psi} = 3.096$~GeV, $M_{\psi'} =
3.686$~GeV, $M_{\chi_{c0}} = 3.415$~GeV, $M_{\chi_{c1}} =
3.510$~GeV, $M_{\chi_{c2}} = 3.556$~GeV, $M_{\eta_c} = 2.984$~GeV.
The branching fractions~\cite{ParticleDataGroup:2020ssz} of
charmonium states into lower-energy states and $J/\psi$ branching
fractions into lepton pairs are Br$(\chi_{c0} \rightarrow J/\psi +
\gamma) = 0.014$, Br$(\chi_{c1} \rightarrow J/\psi + \gamma) =
0.343$, Br$(\chi_{c2} \rightarrow J/\psi + \gamma) = 0.19$,
Br$(\psi' \rightarrow J/\psi + X) = 0.614$, Br$(J/\psi \rightarrow
e^{+} e^{-}) = 0.05971$, Br$(J/\psi \rightarrow \mu^+ \mu^-) =
0.05961$. For color singlet states the following
LDMEs~\cite{Braaten:1999qk} were used: $\langle \mathcal{O}^{J/\psi}
[^3S_1^{(1)}] \rangle = 1.3$~GeV$^3$, $\langle \mathcal{O}^{\psi'}
[^3S_1^{(1)}] \rangle = 0.65$~GeV$^3$, $\langle
\mathcal{O}^{\chi_{c0}} [^3P_0^{(1)}] \rangle = 0.089$~GeV$^5$ and
$\langle \mathcal{O}^{\eta_c} [^1S_0^{(1)}] \rangle =
{0.44}$~GeV$^3$.

We make matching of the CPM and TMD PM contributions for description
of the charmonium production at arbitrary values of charmonium
transverse momentum. The transverse mass of charmonium $M_T^{} =
\sqrt{M^2 + p_T^2}$ in the CPM calculations and the mass of
charmonium $M$ in the SGR approach were used as scales of
factorization, $\mu$, and renormalization, $\mu_{R}^{}$. In order to
calculate the feed-down contributions to the prompt $J/\psi$
production of the kind $\mathcal{C}' \to \mathcal{C} + X$ correctly,
the transverse momentum shift effect was taken into account:
$p_{T\mathcal{C}}^{} \approx (M_\mathcal{C}/M_{\mathcal{C}'}) \cdot
p_{T\mathcal{C}'}^{}$.

\subsection{$\bm{\eta_c}$ production}
\label{eta_c}

First, we will consider the $\eta_c$ production as a clearer
case. Here we provide estimations for the $\eta_c$ production at small
$p_T^{}$ where the color singlet state contribution $^1S_0^{(1)}$ is
the LO of the NRQCD series. That's why, in contrast to the $J/\psi$
production case, there is no need to seek a source of the color
octet LDMEs in order to fit them, which is a generally problematic
task due to the narrow fitting region $p_T^{} \ll M$ in the TMD PM.
In any case, there is no data available for the $\eta_c$
hadroproduction at the small $p_T^{}$ values. For $\eta_c$ we can
only make theoretical predictions for the color singlet contribution
with a known color singlet LDME.

\begin{figure}[h!]
\begin{center}
\begin{minipage}[h]{0.49\linewidth}
\includegraphics[width=6cm]{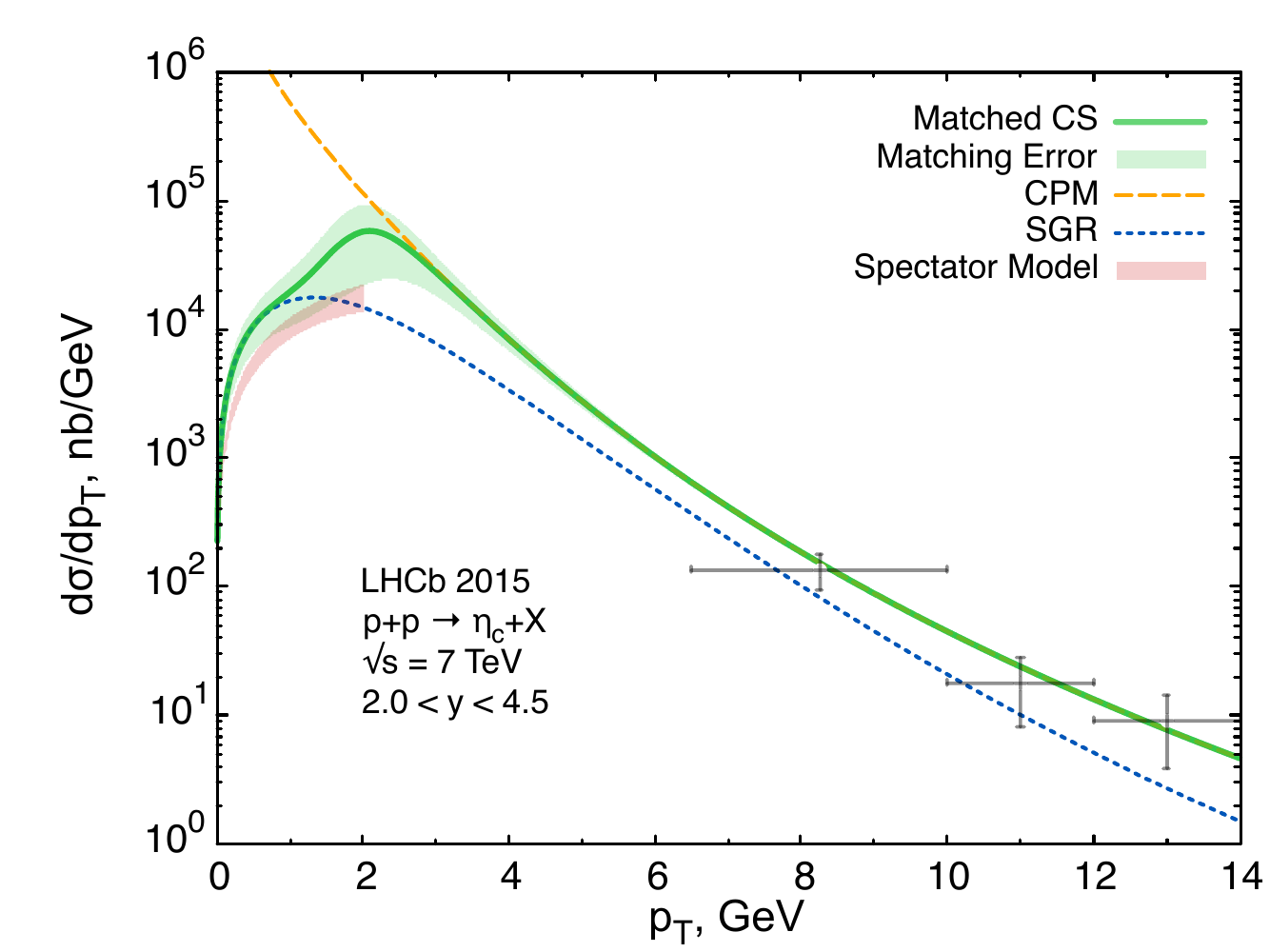}
\end{minipage}
\hfill
\begin{minipage}[h]{0.49\linewidth}
\includegraphics[width=6cm]{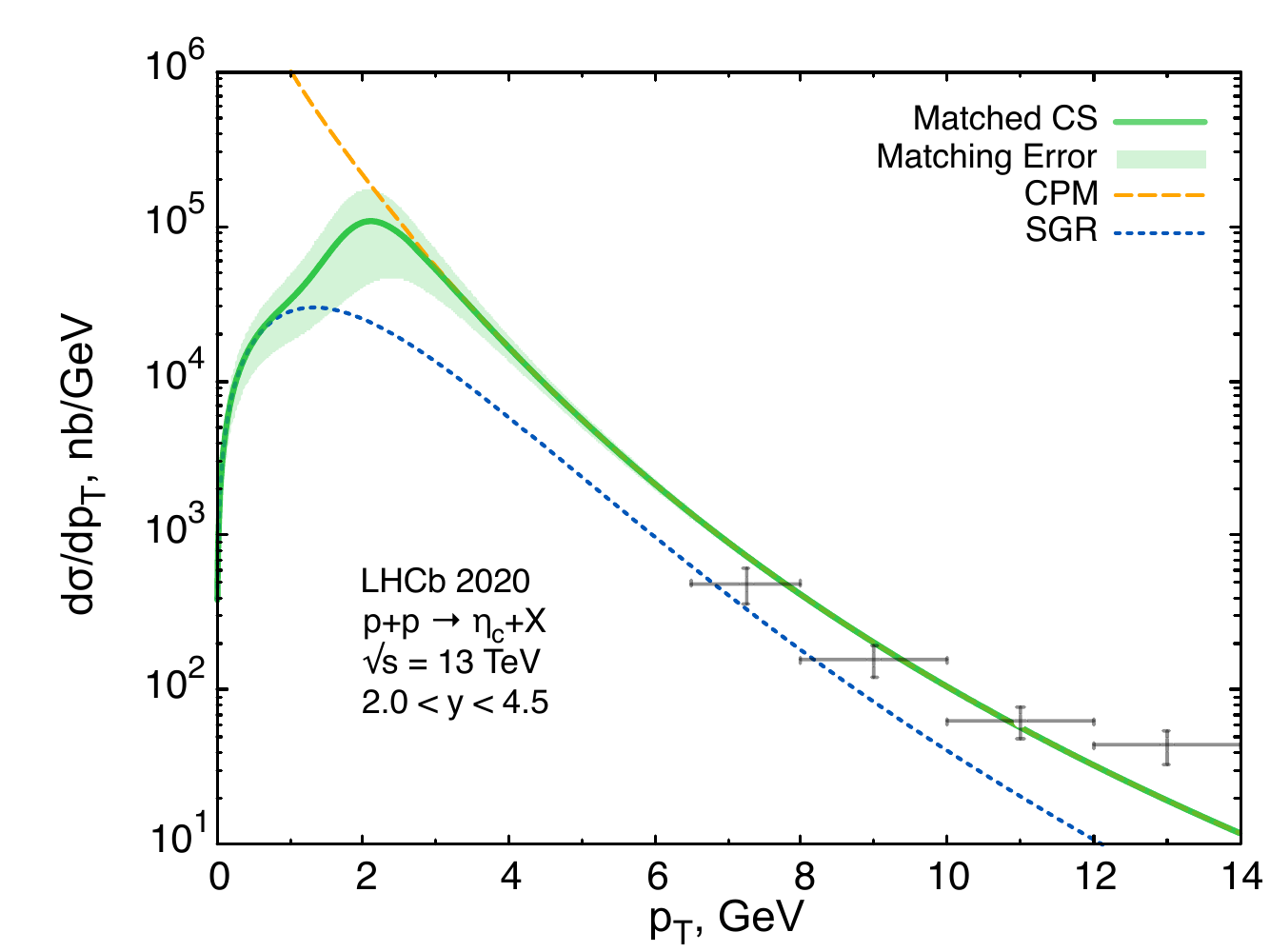}
\end{minipage}
\caption{The cross section of the $\eta_c$ production as a function
of transverse momentum for forward rapidity region $2<y<4.5$ at the
energies $\sqrt{s} = 7$ TeV (left panel)  and $13$ TeV (right
panel). The spectator model prediction is from Ref.
\cite{Bacchetta:2022nyv}. The data are from the LHCb
Collaboration~\cite{LHCb:2014oii,LHCb:2019zaj}} \label{fig:1}
\end{center}
\end{figure}

The predictions for the $\eta_c$ production cross section in the
LL-LO accuracy are shown in Fig.~\ref{fig:1} for the LHCb
 {energies} of $\sqrt{s} = 7$ and $13$
TeV~\cite{LHCb:2019zaj, LHCb:2014oii}. At first, it is interesting
that the LO CPM using the CSM calculation does not contradict LHCb
data with respect to the NLO CPM using CSM
calculation~\cite{Butenschoen:2014dra}, which overestimates LHCb
data a little when the value $\langle \mathcal{O}^{\eta_c}
[^1S_0^{(1)}] \rangle =\frac{1}{3} \langle \mathcal{O}^{J/\psi}
[^3S_1^{(1)}] \rangle \simeq   {0.44}$~GeV$^3$ is used. The next
surprising finding is a good agreement of the SGR calculation with
the data up to $p_T \simeq 14$ GeV. It looks as an artifact of the
SGR approach, which should only be applicable in the region of small
$p_T< M_{\eta_c}$. In any case, the matched cross section in this
region of the transverse momenta $p_T$ is defined by the CPM
contribution only.

\begin{figure}[h!]
\begin{center}
\begin{minipage}[h]{0.49\linewidth}
\includegraphics[width=6cm]{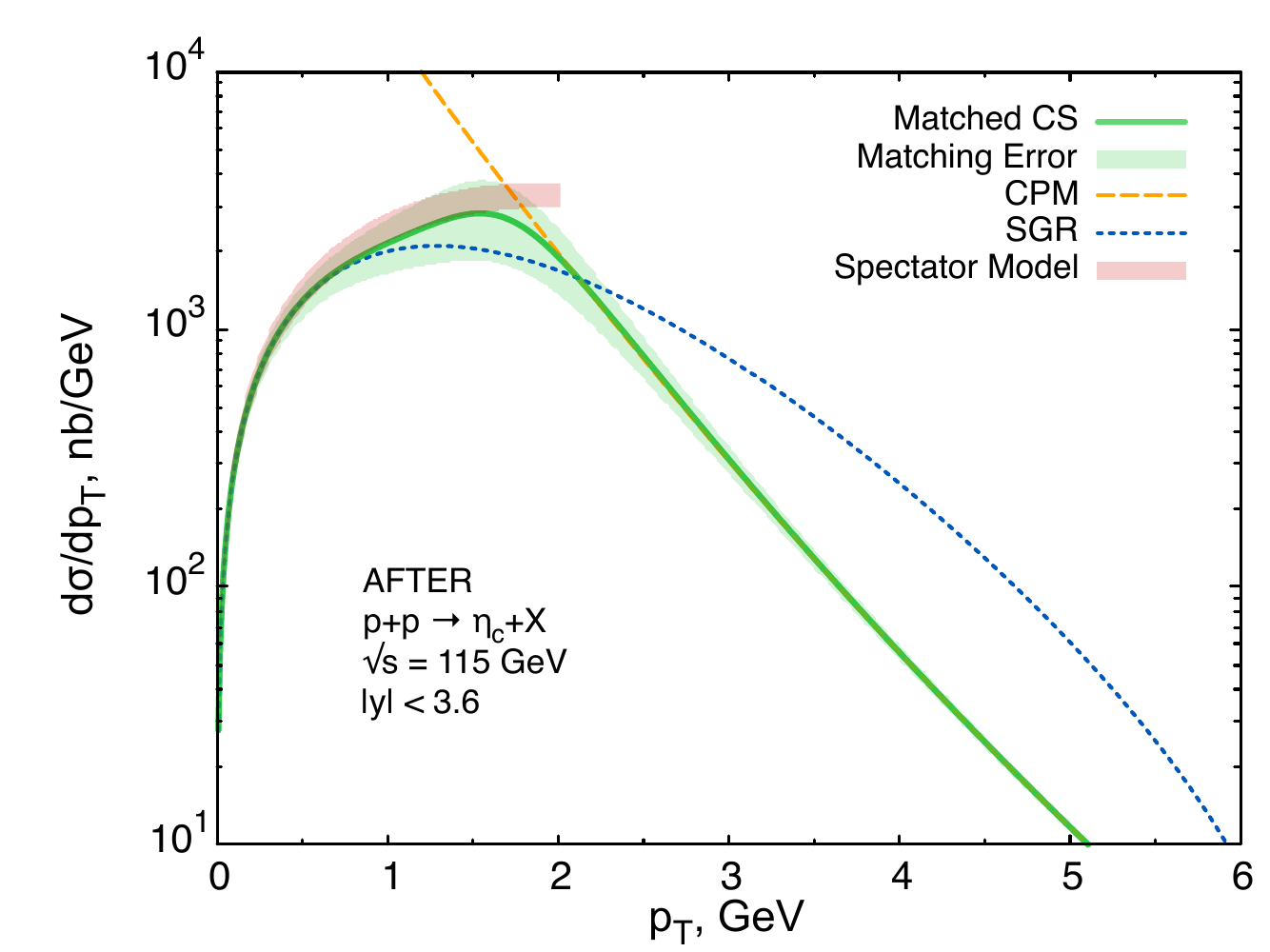}
\end{minipage}
\hfill
\begin{minipage}[h]{0.49\linewidth}
\includegraphics[width=6cm]{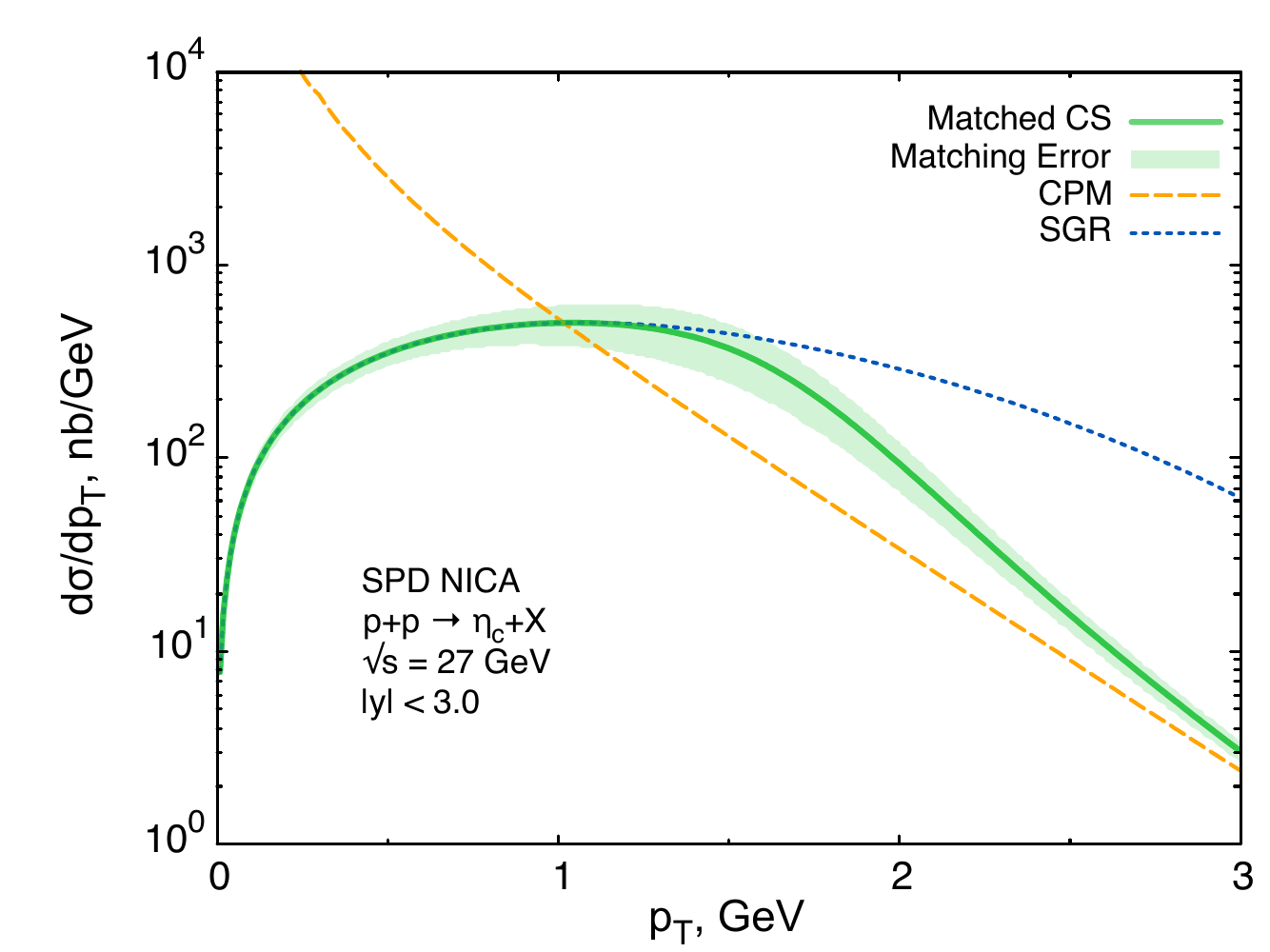}
\end{minipage}
\caption{The cross section of the $\eta_c$ production as a function
of transverse momentum at the energies $\sqrt{s} = 115$ GeV (left
panel) and $\sqrt{s} = 27$ GeV (right panel).  The spectator model
prediction is from Ref. \cite{Bacchetta:2022nyv}.}\label{fig:2}
\end{center}
\end{figure}

In Fig.~\ref{fig:2}, our predictions for the $\eta_c$ production differential
cross sections within the matched SGR plus CPM approach at the
$\sqrt{s} = 27$ GeV for the future SPD NICA~\cite{Arbuzov:2020cqg}
and at the $\sqrt{s} = 115$ GeV for the AFTER@LHC
experiments~\cite{Lansberg:2012sq} are presented.

{In Figs. \ref{fig:1} and \ref{fig:2}, the predictions obtained in
the TMD approach using the spectator model~\cite{Bacchetta:2022nyv}
for gluon PDF at the low initial scale are also presented to compare
at the region of small transverse momenta $p_T < 2$ GeV. The
unpolarized gluon TMD
 {PDF} and the Boer-Mulders one were used in terms of
the PVGlueModel20 model of Ref. \cite{Bacchetta:2020vty} at the
initial scale. Their evolution {is} controlled by the standard
Collins-Soper-Sterman equations \cite{Collins:2011zzd}. We find
rather good agreement at the energy $\sqrt{s}=115$ GeV and our
prediction sufficiently overestimates  {the} result of
Ref.~\cite{Bacchetta:2022nyv} at the energy $\sqrt{s}=7$ TeV. Note,
{according to} Ref.~\cite{Bacchetta:2022nyv}, the contribution of
the Boer-Mulders function in the $\eta_c$ production cross sections
is about a few percent and {the} one may be neglected  in our
predictions.}

\subsection{$\bm{J/\psi}$ production}
\label{Jpsi}

The $J/\psi$ production is a more sophisticated sample for test of
factorization approaches.  {As opposed to} the $\eta_c$
hadroproduction at the small transverse momenta, when the
gluon-gluon fusion is absolutely dominant contribution, the
quark-antiquark annihilation contribution to the $J/\psi$ production
cannot be neglected and it is estimated to be about $10\%$ of the
total cross section at the $\sqrt{s} = 200$ GeV and it is about
$20-30 \%$ at smaller energies~\cite{Chernyshev:2022gek}. In this
paper, we take into consideration quark-antiquark annihilation
subprocesses  {both} in the SGR approach  {and} in the CPM
calculation.

We do not take into account the contribution of the octet states $\psi'$
to the prompt production of $J/\psi$ because of their very small value
compared to the contributions of the analogous states of $J/\psi$
and their almost identical dependence on $p_T^{}$ in the LDME
fitting in the available domains, since both contributions are
described by the same hard amplitudes. One can consider the small
octet contributions $\psi'$ to be effectively included in the direct
production of the corresponding $J/\psi$ states.

The fit of the octet LDMEs was performed with the experimental data sets
of the PHENIX~\cite{PHENIX:2011gyb, PHENIX:2006aub} and
STAR~\cite{STAR:2018smh, STAR:2009irl} collaborations for the
$J/\psi$ production in proton--proton collisions at $\sqrt{s} = 200$
GeV in different rapidity intervals, commonly in the SRG ($p_T^{} <
1$~GeV) and in the CPM ($p_T^{} > 5$ GeV) applicability domains
under assumption of independence of the LDMEs from the factorization
model. The identical dependence on $p_T^{}$ of the $^1S_0^{(8)}$ and
$^3P_J^{(8)}$ contributions to the $J/\psi$ direct production does
not allow us to separately extract the corresponding LDMEs within
the only one factorization model, so their values are usually
obtained only as a linear combination, however, the common fit in
the CPM and in the SGR, where the LDMEs are included in two
different linear combinations, makes it possible to find both
$\langle \mathcal{O}^{J/\psi}[^1S_0^{(8)}] \rangle$ and $\langle
\mathcal{O}^{J/\psi}[^3P_0^{(8)}] \rangle$ separately. In addition,
the relations between the LDMEs due to the heavy quark spin symmetry in
the leading $\upsilon$ order of the NRQCD were used: $ \langle
\mathcal{O}^{\cal C}[^3P_J^{(8)}] \rangle = (2J+1) \cdot \langle
\mathcal{O}^{\cal C}[^3P_0^{(8)}] \rangle$.

The values of the octet LDMEs obtained by fitting the experimental
data are in Table~\ref{ta1}, where the uncertainties correspond to
one standard deviation. The ME for the $^3S_1^{(8)}$ states of
$J/\psi$ and $\chi_{cJ}$ is obviously the same and that follows {a}
zero value of the LDME and {an} inability of the corresponding LDME
extraction due to the close $p_T^{}$-dependence of these states'
cross section up to $p_T^{} \simeq 9$ GeV. More data for large
$p_T^{}$ at not very high energies may provide an opportunity to
separate them.

\begin{figure}[h!]
\begin{center}
\begin{minipage}[h]{0.49\linewidth}
\includegraphics[width=6cm]{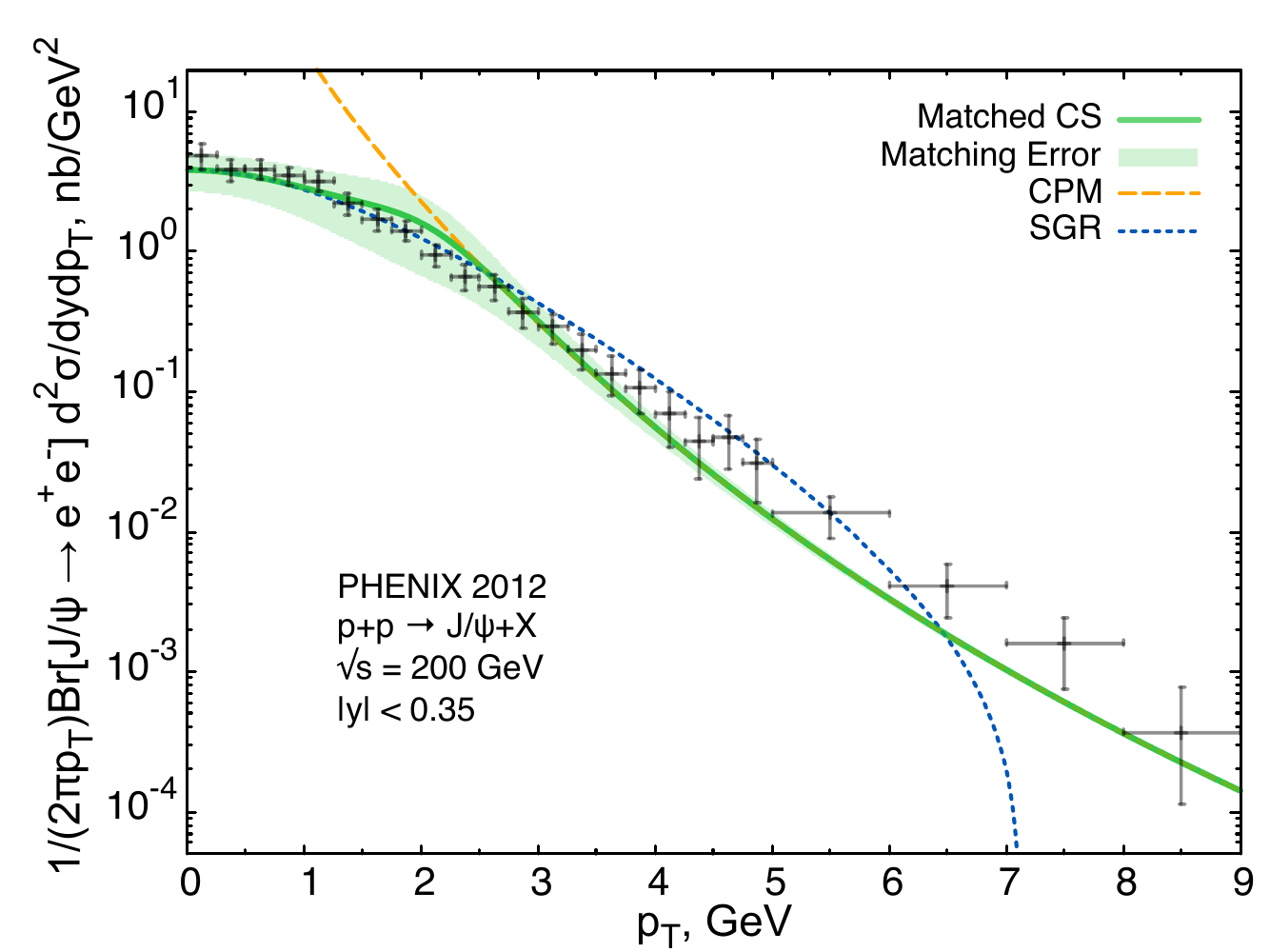}
\end{minipage}
\hfill
\begin{minipage}[h]{0.49\linewidth}
\includegraphics[width=6cm]{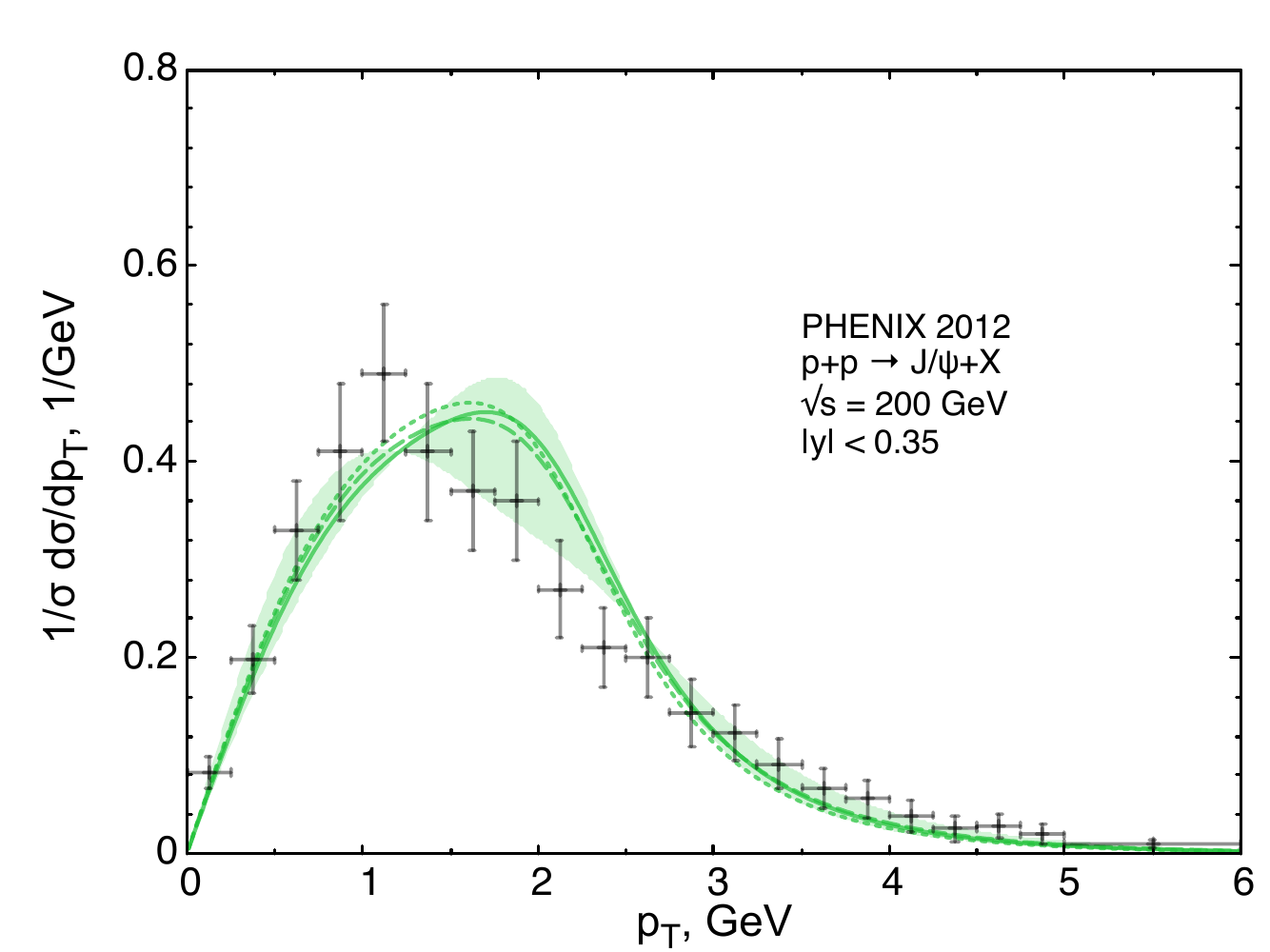}
\end{minipage}
\caption{The cross section of the $J/\psi$ production as a function
of transverse momentum at the energy $\sqrt{s} = 200$ GeV and
$|y|<0.35$ (left panel), and the corresponding normalized spectrum
(right panel). The data are from the PHENIX
Collaboration~\cite{PHENIX:2011gyb}.}\label{fig:3}
\end{center}
\end{figure}

\begin{figure}[h!]
\begin{center}
\begin{minipage}[h]{0.49\linewidth}
\includegraphics[width=6cm]{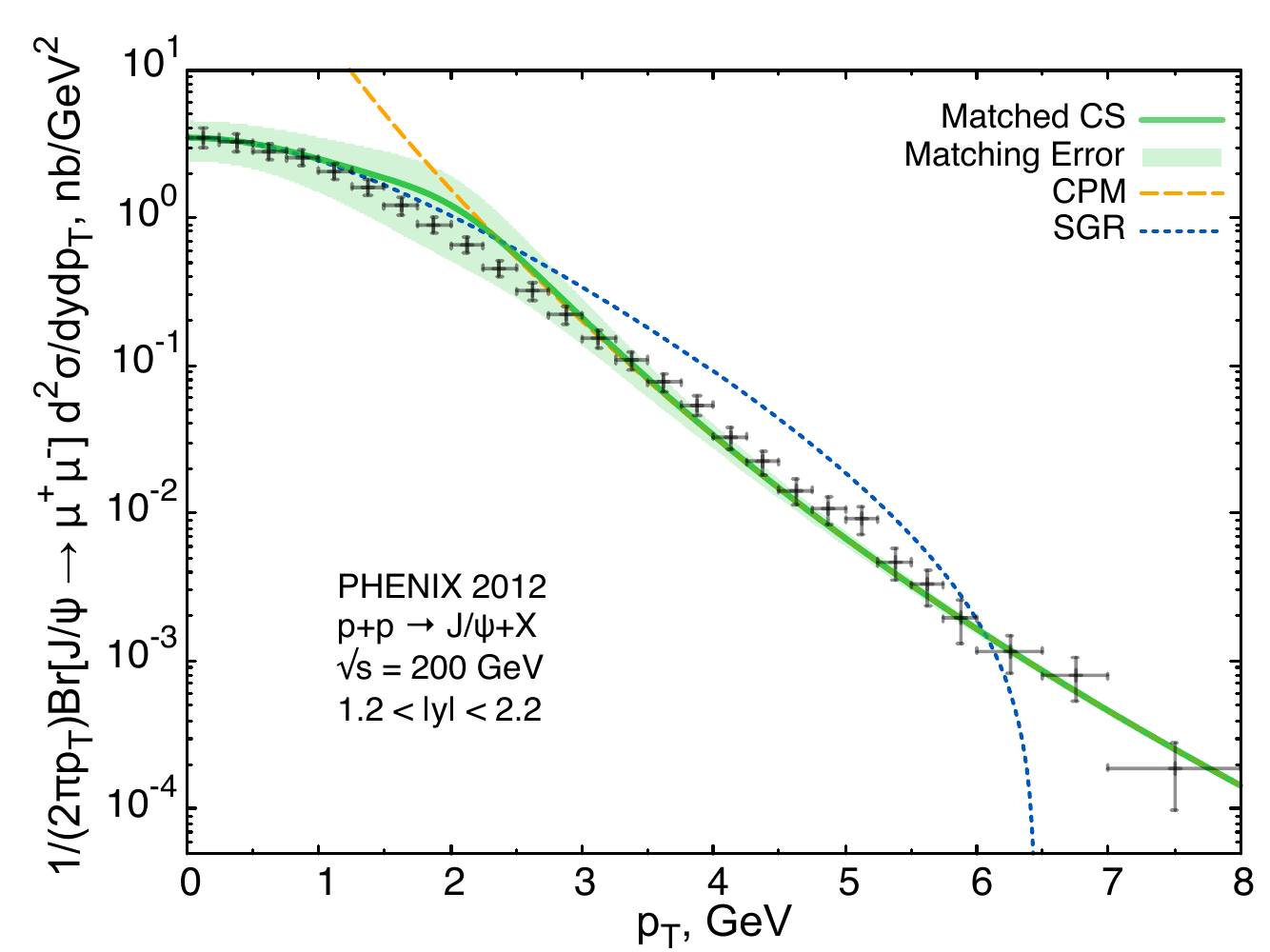}
\end{minipage}
\hfill
\begin{minipage}[h]{0.49\linewidth}
\includegraphics[width=6cm]{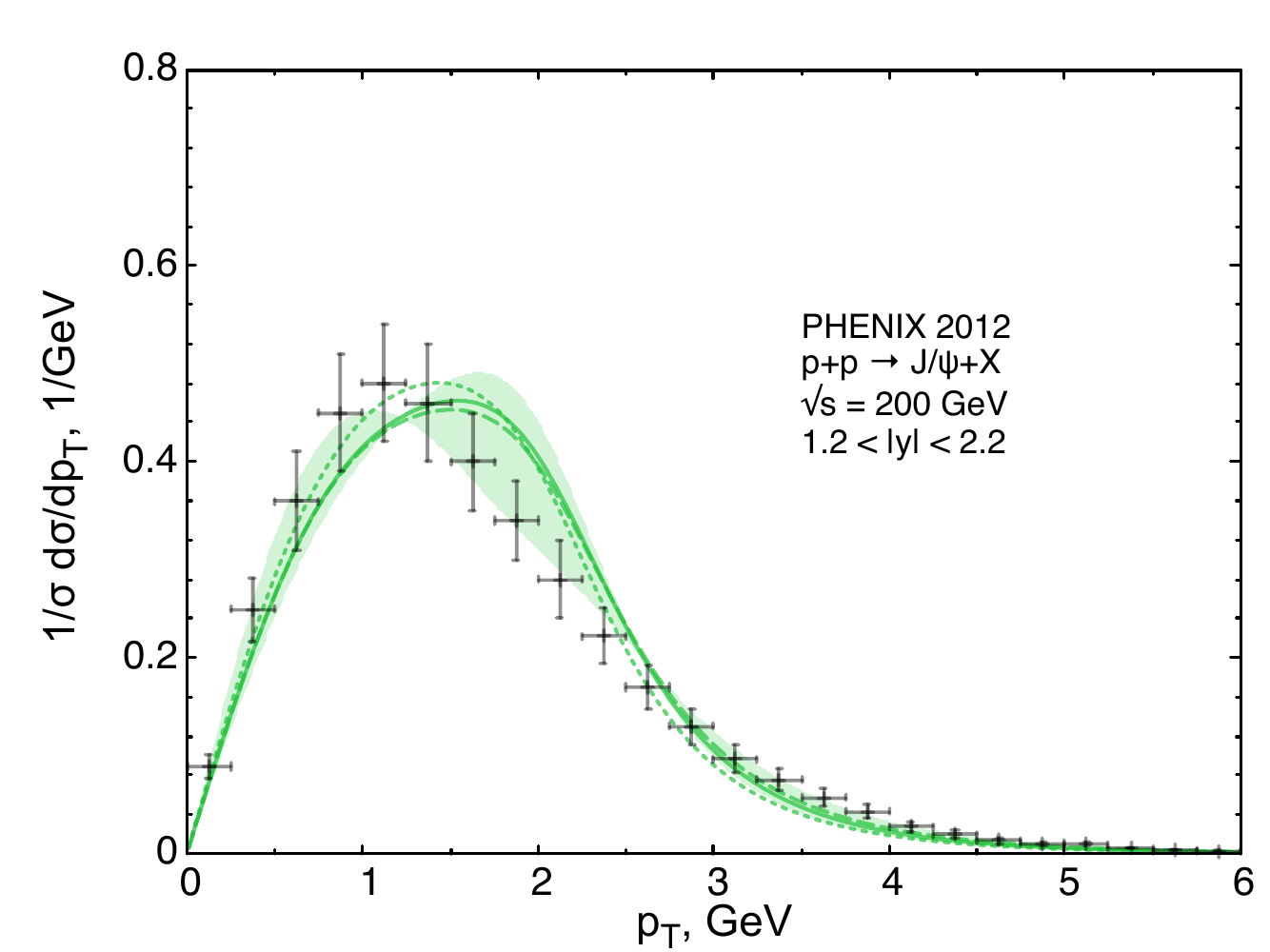}
\end{minipage}
\caption{The cross section of the $J/\psi$ production versus
transverse momentum at the PHENIX energy of $\sqrt{s} = 200$ GeV and
$1.2<|y|<2.2$ (left panel), and the corresponding normalized
spectrum (right panel). The data are from the PHENIX
Collaboration~\cite{PHENIX:2006aub}.}\label{fig:4}
\end{center}
\end{figure}

\begin{figure}[h!]
\begin{center}
\begin{minipage}[h]{0.49\linewidth}
\includegraphics[width=6cm]{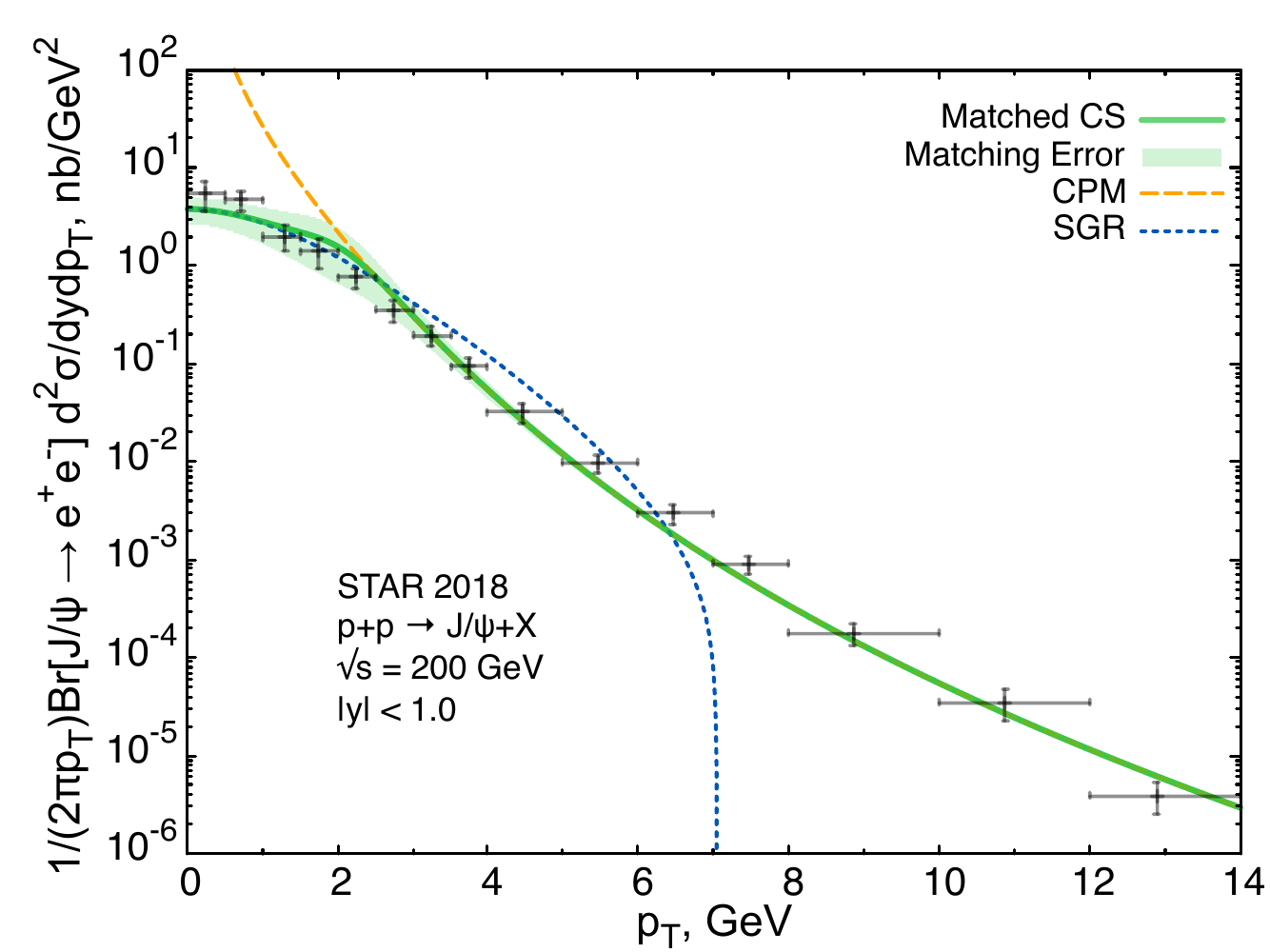}
\end{minipage}
\hfill
\begin{minipage}[h]{0.49\linewidth}
\includegraphics[width=6cm]{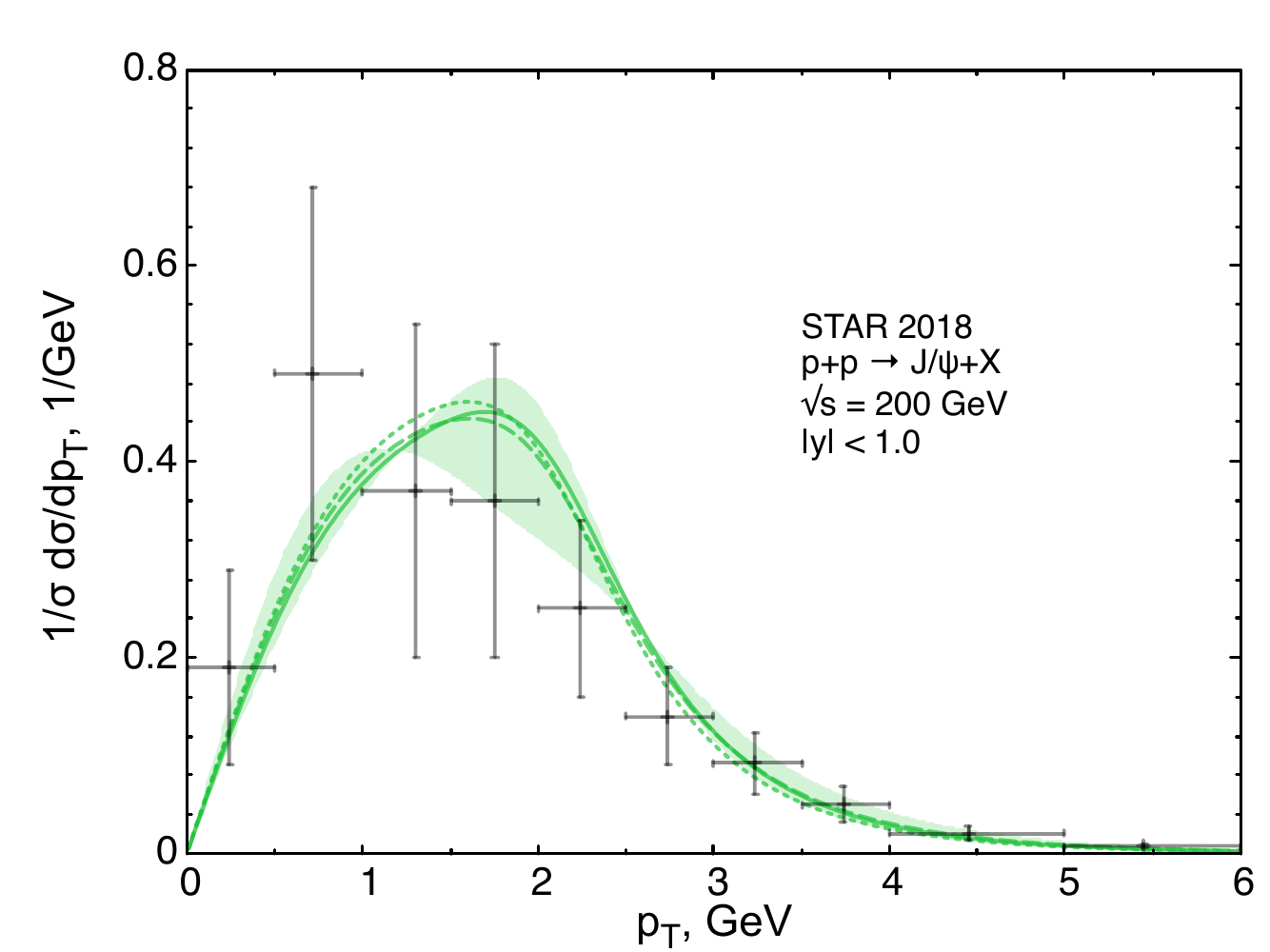}
\end{minipage}
\caption{The cross section of the $J/\psi$ production as a function
of transverse momentum at the energy  $\sqrt{s} = 200$ GeV and
$|y|<1.0$ (left panel), and the corresponding normalized spectrum
(right panel). The data are from the STAR
Collaboration~\cite{STAR:2018smh,STAR:2009irl}}\label{fig:5}
\end{center}
\end{figure}

The results of cross section calculations for the kinematics of
the PHENIX~\cite{PHENIX:2011gyb} and STAR~\cite{STAR:2018smh}
experiments are plotted in Figs.~\ref{fig:3}, \ref{fig:4}, \ref{fig:5} together with their normalized
spectra. Theoretical calculations for the 2007 PHENIX experimental
data set~\cite{PHENIX:2006aub} are not shown here, since they fully
coincide with those in Fig.~\ref{fig:3} and \ref{fig:4}. The light green bands in the
plots show the uncertainties of the cross section within the
matching procedure.

\begin{table}[h!]
\tbl{Result for LDMEs fitting on $J/\psi$ production within domains of $p_T^{} < 1$ GeV (SGR) and $p_T^{} > 5$ GeV (CPM).}
{\begin{tabular}{|l|c|}
\hline
$\langle \mathcal{O}^{J/\psi}[^{1}S_0^{(8)}] \rangle$, GeV$^{3}$ & $(8.73 \pm 0.40) \cdot 10^{-2}$ \\ \hline
$\langle \mathcal{O}^{J/\psi}[^{3}P_0^{(8)}] \rangle$, GeV$^{5}$ & $(2.35 \pm 1.48) \cdot 10^{-3}$ \\ \hline
$\langle \mathcal{O}^{J/\psi}[^{3}S_1^{(8)}] \rangle$, GeV$^{3}$ & $(7.33 \pm 1.22) \cdot 10^{-3}$  \\ \hline
$\langle \mathcal{O}^{\chi_{c0}}[^{3}S_1^{(8)}] \rangle$, GeV$^{3}$ & $(0.00 \pm 1.63) \cdot 10^{-3}$  \\ \hline
$\chi^2/\textnormal{n.d.f.}$ & $1.38$ \\ \hline
\end{tabular}\label{ta1} }
\end{table}

Additional uncertainties in the cross section predictions are
introduced by the choice of a hard scale and the LDME variations.
However, these uncertainties almost disappear when the
normalized spectra are calculated. The plots below show, in addition
to the main theoretically predicted differential cross sections,
also the curves while varying different scales by a factor of 2. The
LDME uncertainty hardly influence on the shape of the spectrum
curves, so it cannot be seen on the plots.

Our calculations show that the contribution of the quark-antiquark
annihilation subprocesses for the central and middle intervals of
rapidity at $\sqrt{s} = 200$ GeV is about $8\%$, and the feed-down
contribution from the  $\chi_{cJ}^{} \rightarrow J/\psi + \gamma$
decays is only about $4\%$, while the experimental estimation for
the $P$-wave contributions is about $30\%$. This contradiction requires a special investigation.

\begin{figure}[h!]
\begin{center}
\begin{minipage}[h]{0.49\linewidth}
\includegraphics[width=6cm]{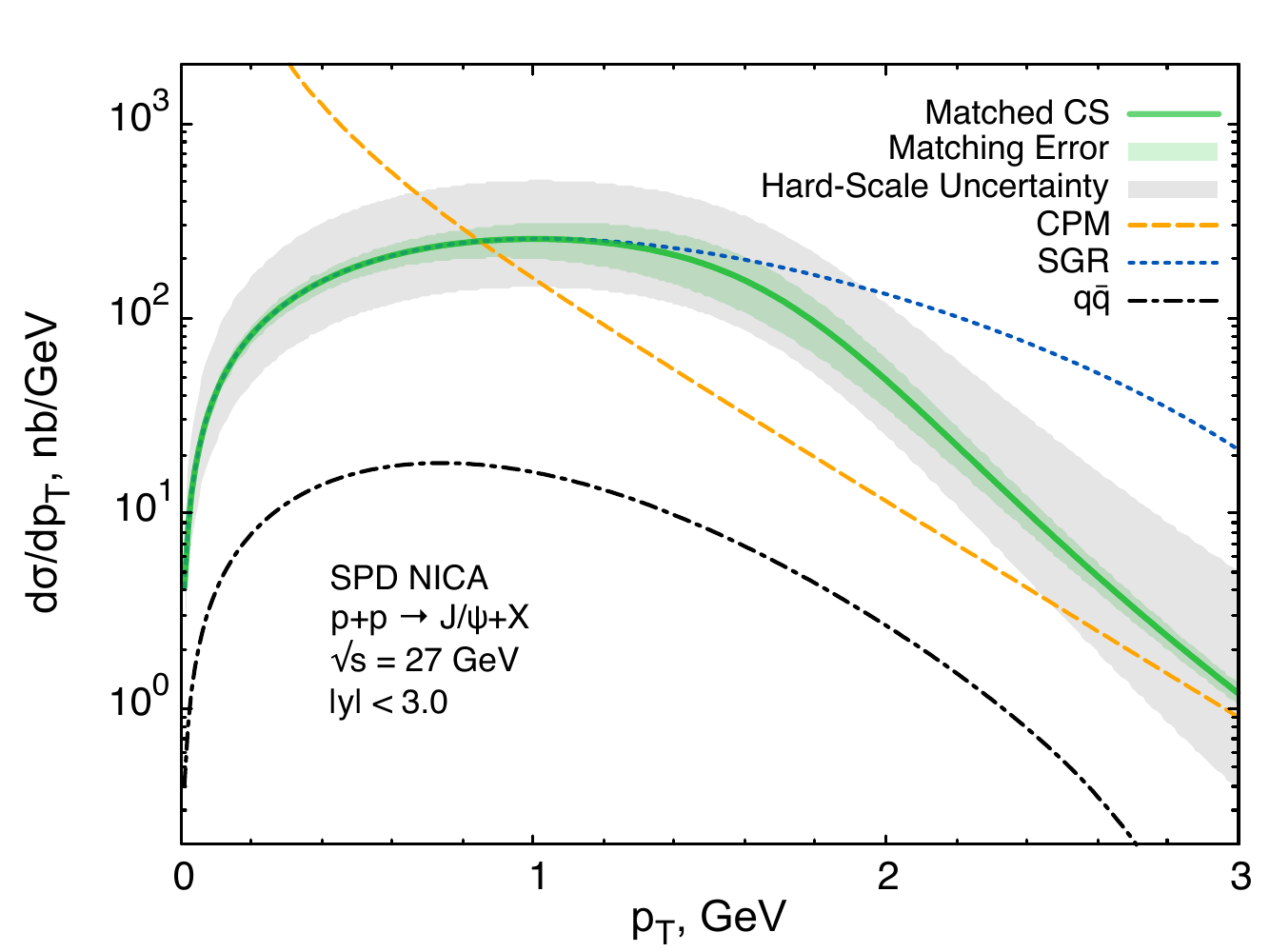}
\end{minipage}
\hfill
\begin{minipage}[h]{0.49\linewidth}
\includegraphics[width=6cm]{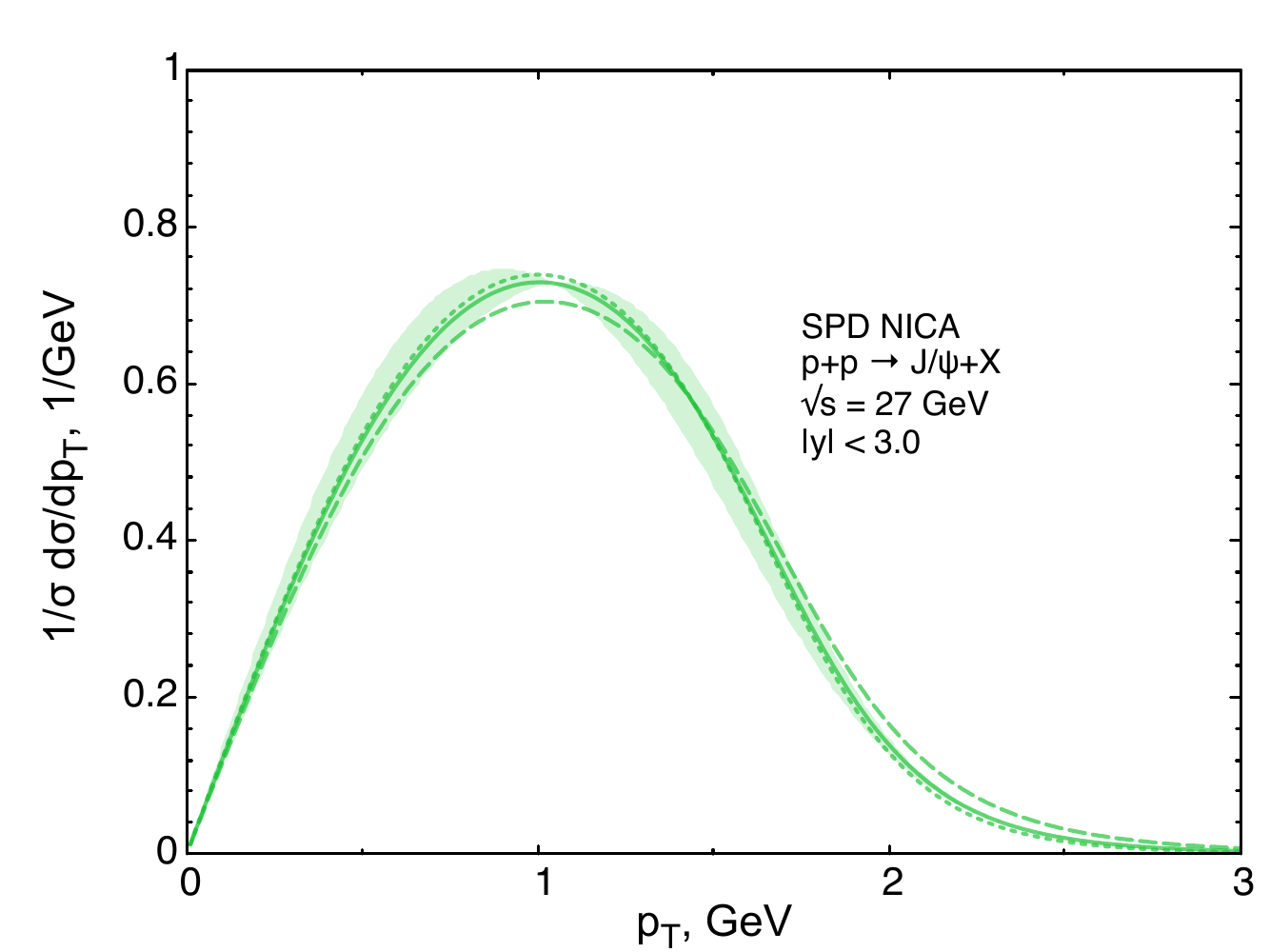}
\end{minipage}
\caption{The cross section of the $J/\psi$ production as a function
of transverse momentum at the NICA energy of $\sqrt{s} = 27$ GeV
(left panel), and the corresponding normalized spectrum (right
panel).}\label{fig:6}
\end{center}
\end{figure}

We present predictions for the $J/\psi$ production in the SPD NICA
kinematics at $\sqrt{s} = 27$ GeV with the fitted LDMEs. As can be
seen in Fig.~\ref{fig:6}, as the energy $\sqrt{s}$ decreases, the
transition region from one factorization to another shifts towards
the smaller values of $p_T^{}$. The estimated contribution of
quark-antiquark subprocesses is less than $10\%$, the fraction of
$\chi_{cJ}^{}$ decays is about $5\%$.

Theoretical predictions based on different factorization approaches
and different hadronization models are very sensitive to the
production of the polarized of $J/\psi$ mesons. It refers to the so-called "quarkonium polarization puzzle", which implies a discrepancy
between experimental data and current predictions of the polarized
$J/\psi$ production cross sections. At the considered center-of-mass
energy, we can analyze the $J/\psi$ polarization
data~\cite{PHENIX:2009ghc} of the PHENIX collaboration for $p_T^{}
\leq 5$ GeV to compare with. The data are provided in the helicity
frame, where the axis of the longitudinal polarization is directed
along the three-dimensional momentum of the $J/\psi$. To calculate
polarized direct $J/\psi$ cross section we should only know that
the vector of $J/\psi$ longitudinal polarization in the helicity
frame of the $J/\psi$ can be written in the following
way~\cite{Beneke:1998re}:
\begin{equation}
\varepsilon_\mu^{}(J_z = 0) = \frac{(p Q)p_{\mu}^{}/M - M
Q_{\mu}^{}}{\sqrt{(p Q)^2 - s M^2}},
\end{equation}
where $p$ is the $J/\psi$ 4-momentum, $Q$ is the sum of the initial
proton's 4-momenta and $s = Q^2$. To calculate the feed-down
contribution in the prompt polarized $J/\psi$ production cross
section, we need {the} known polarization tensors for the $P$-wave
states, which can be obtained by summing the products of the vectors
$\varepsilon_\mu^{} (J_z)$ and the corresponding Clebsch--Gordan
coefficients:
\begin{equation}
\varepsilon_{\mu\nu}^{(J)}(J_z^{}) = \sum\limits_{J_{1z}, J_{2z}} C_{J_1 J_{1z}, J_2 J_{2z}}^{J J_z} \varepsilon_\mu(J_{1z})\varepsilon_\nu(J_{2z}).
\end{equation}
The explicit NRQCD scheme and relevant formulae for calculation of polarized $J/\psi$ cross section due to the feed-down decays of
$\chi_{cJ}$ and $\psi'$ can be found in Ref.~\cite{Kniehl:2000nn}.

\begin{figure}[h!]
\begin{center}
\begin{minipage}[h]{0.49\linewidth}
\includegraphics[width=6cm]{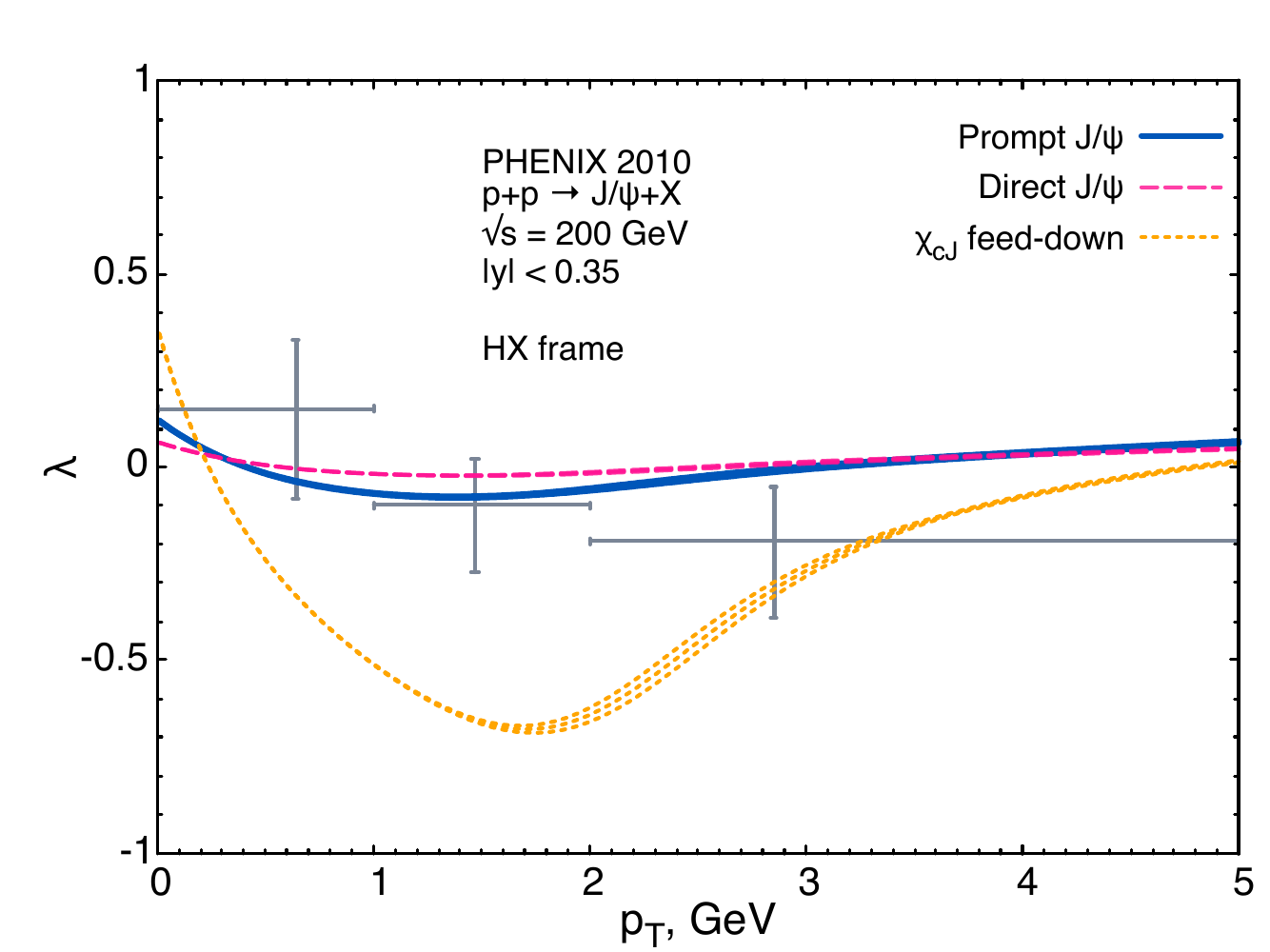}
\end{minipage}
\hfill
\begin{minipage}[h]{0.49\linewidth}
\includegraphics[width=6cm]{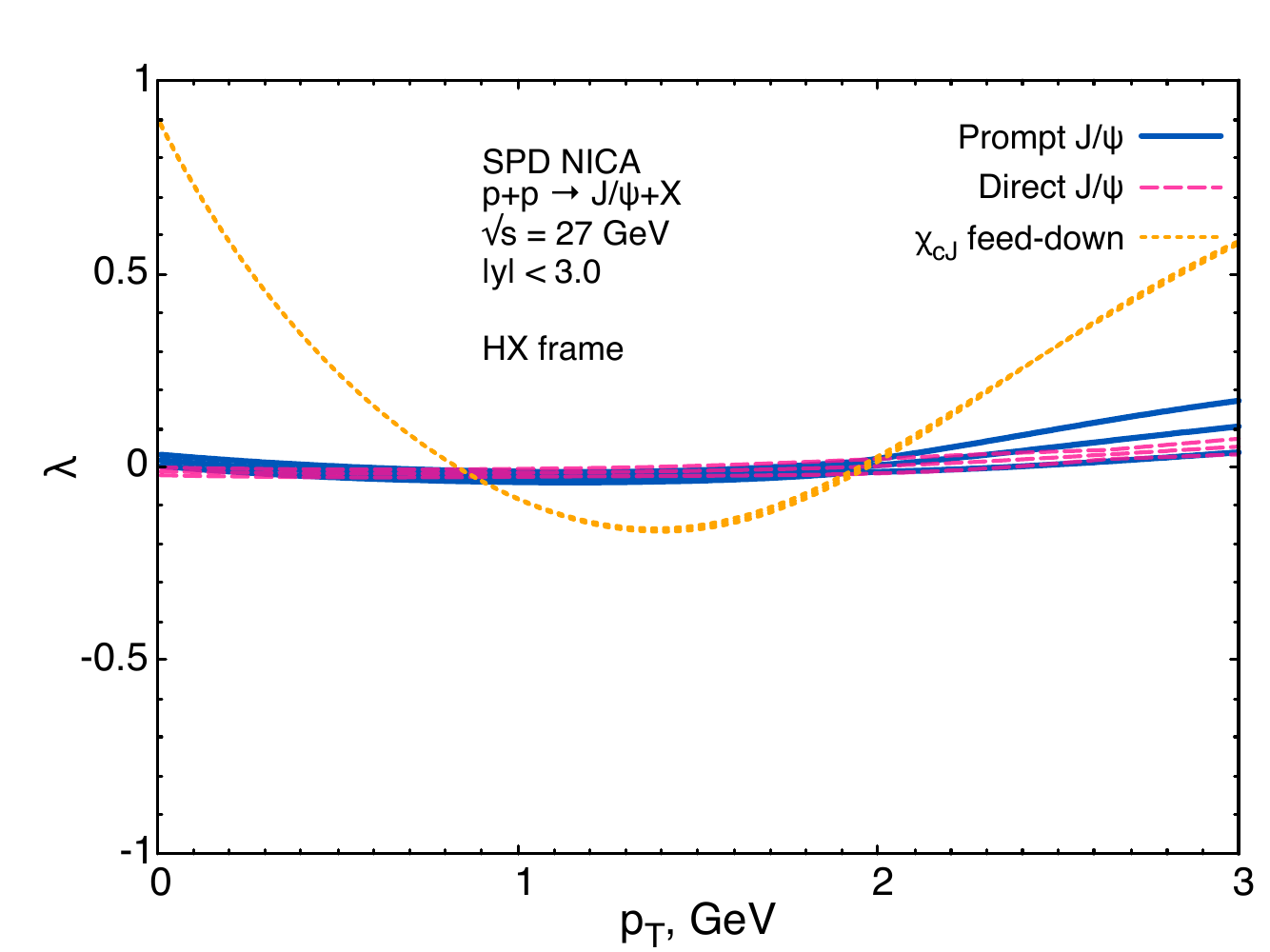}
\end{minipage}
\caption{The angular coefficient $\lambda$ for $J/\psi$ production
within the helicity frame as a function of transverse momentum at
the energy $\sqrt{s} = 200$ GeV, $|y|<0.35$ for PHENIX (left panel)
and at the energy $\sqrt{s} = 27$ GeV, $|y|<3.0$ for SPD NICA (right
panel). The data are from the PHENIX
Collaboration~\cite{PHENIX:2009ghc}.}\label{fig:7}
\end{center}
\end{figure}

We calculate the coefficient $\lambda$, which is one of the angular
coefficients in the lepton angular distribution in $J/\psi\to l\bar
l$ decay. This coefficient is expressed in terms of the polarized
($L$ for longitudinal and $T$ for transverse polarization) cross
sections:
\begin{equation}
\lambda = (\sigma_{T} - 2\sigma_{L})/(\sigma_{T} + 2\sigma_{L}).
\end{equation}
The comparison of the experimental data with the calculation results is
given in Fig.~\ref{fig:7}, where the channels of direct $J/\psi$ production
and $\chi_{cJ}$ states decays are shown separately as well. We find
good agreement of the calculation with the data at small~$p_T$, the
application domain of the TMD factorization theorem, and
disagreement within the intermediate $p_T^{}\sim M$ region where the
prediction is made using the matching InEW scheme. The polarization
of the $J/\psi$ is described with the direct production
predominantly, though the channel of the $P$-wave states decays
could describe polarization at $p_T^{}> M$. This may be an evidence for
biased fitting of LDMEs for $\chi_{cJ}$, so a more precise description
of the $\chi_{cJ}$ production could fix both the prediction of
polarization at intermediate $p_T^{}$ and the problem of $P$-wave
states fraction in unpolarized $J/\psi$ production which was
mentioned above.

\section{Conclusion}

In the present paper, we studied the prompt production of $S$-wave
states of charmonium, these are $J/\psi$ and $\eta_c$, within the
SGR approach as a TMD PM factorization framework. This TMD PM allows
to describe the charmonium production cross section at small $p_T^{}$
and provides the description of the transverse momentum spectrum at
moderate and large $p_T^{}$ within the fixed-order CPM calculation
using the InEW scheme for matching the SGR approach and CPM.

Firstly, we provide calculations for $\eta_c$ production within the
CSM at the LL-LO accuracy of the SGR approach matched with the LO
CPM calculations for energies of present and future experiments:
LHCb at $\sqrt{s} = 13$ TeV, AFTER@LHC at $\sqrt{s} = 115$ GeV and
SPD NICA at $\sqrt{s} = 27$ GeV. Secondly, we calculate the prompt
$J/\psi$ production cross sections at the energy $\sqrt{s} = 200$
GeV (PHENIX, STAR) and at $\sqrt{s} = 27$~GeV (SPD NICA) at LL-LO
approximation of the SGR, matched with the LO CPM calculation using
the NRQCD. We obtained the color octet LDMEs by fitting them to the
experimental data at the $\sqrt{s} = 200$~GeV and used them for
prediction of cross section for the SPD NICA energy. We estimate
feed-down fraction as well as the relative role of the gluon-gluon
fusion and the quark-antiquark annihilation subprocesses in the
prompt $J/\psi$ production at the different energies. Thirdly, we
describe the PHENIX data for polarized prompt $J/\psi$ production in
the helicity frame at the energy $\sqrt{s}=200$ GeV and $p_T<M$, and
make prediction for the future SPD NICA experiment. For both
kinematic conditions, $J/\psi$ are predicted to be nearly
unpolarized for all $p_T^{}$ value, although this result is in agreement with
the PHENIX data at small $p_T^{}$.

Thus, we demonstrate the relevance of the combined scheme based on the
matching between the SGR approach and the fixed-order CPM for
the calculation of $J/\psi$ and $\eta_c$ production cross sections. We
should note that the calculation in the NLL-NLO approximation of the
SGR approach and at the NLO level of the fixed-order CPM look more
accurate and should be made in the future. However, the LDMEs used in
the calculations should be redefined appropriately to describe
the data. After this, the new theoretical predictions should be very
close to those obtained within the present approximations.

\section{Acknowledgments}

We are grateful to Alexey Guskov, Oleg Teryaev and other members of
the SPD NICA Collaboration for a fruitful discussion of the obtained
results. The work is supported by the Foundation for the Advancement
of Theoretical Physics and Mathematics BASIS, grant No. 24-1-1-16-5,
and by the grant of the Ministry of Science and Higher Education of
Russian Federation, No. FSSS-2025-0003.

\bibliographystyle{ws-mpla}
\bibliography{references_9}

\end{document}